\documentclass[aps,twocolumn,showpacs]{revtex4}
\usepackage{epsf}

\begin{document}

\title{\null\vspace*{-1truecm}\hfill\mbox{\small quant-ph/0401154}\\
            \vspace*{0.2truecm}
Optimal Database Search: Waves and Catalysis}
\author{Apoorva Patel}
\affiliation{Centre for High Energy Physics,
             Indian Institute of Science, Bangalore-560012, India}
\email{E-mail: adpatel@cts.iisc.ernet.in}

\begin{abstract}
Grover's database search algorithm, although discovered in the context
of quantum computation, can be implemented using any system that allows
superposition of states. A physical realization of this algorithm is
described using coupled simple harmonic oscillators, which can be exactly
solved in both classical and quantum domains. Classical wave algorithms
are far more stable against decoherence compared to their quantum
counterparts. In addition to providing convenient demonstration models,
they may have a role in practical situations, such as catalysis.
\end{abstract}
\pacs{03.67.Lx, 34.10.+x, 46.40.-f}
\maketitle

\section{The Optimal Search Algorithm}

Database search is an elementary computational task with wide-ranging
applications. Its efficiency is measured in terms of the number of
queries one has to make to the database in order to find the desired item.
In the conventional formulation of the problem, the query is a binary oracle
(i.e. a YES/NO question). For an unsorted database of $N$ items, using
classical Boolean logic, one requires on the average $\langle Q\rangle=N$
binary queries to locate the desired item. The number of queries is reduced
to $\langle Q\rangle=(N+1)/2$, if the search process has a memory so that
an item rejected once is not picked up again for inspection.

Grover discovered a search algorithm that, using superposition of states,
reduces the number of required queries to $Q=O(\sqrt{N})$ \cite{grover}.
This algorithm starts with a superposition state, where each item has an
equal probability to get picked, and evolves it to a target state where
only the desired item can get picked. Following Dirac's notation, and using
the index $i$ to label the items, the starting and target state satisfy
\begin{equation}
|\langle i|s \rangle|^2 = 1/N ~,~~ |\langle i|t \rangle|^2 = \delta_{it} ~~.
\end{equation}
The algorithm evolves $|s\rangle$ towards $|t\rangle$, by discrete rotations
in the two-dimensional space formed by $|s\rangle$ and $|t\rangle$, using
the two reflection operators,
\begin{equation}
U_t = 1 - 2|t\rangle\langle t| ~,~~ U_s = 1 - 2|s\rangle\langle s| ~~,
\end{equation}
\begin{equation}
(-U_sU_t)^Q |s\rangle = |t\rangle ~~.
\end{equation}
$U_t$ is the binary oracle which flips the sign of target state amplitude,
while $-U_s$ performs the reflection-in-the-mean operation.
Solution to Eq.(3) determines the number of queries as
\begin{equation}
(2Q+1) \sin^{-1} (1/\sqrt{N}) = \pi/2 ~~.
\end{equation}
(In practice, $Q$ must be an integer, while Eq.(4) may not have an integer
solution. In such cases, the algorithm is stopped when the state has evolved
sufficiently close to, although not exactly equal to, $|t\rangle$. Then one
finds the desired item with a high probability.)

In the qubit implementation of the algorithm, one chooses $N=2^n$
and the items in the database are labeled with binary digits.
Using a uniform superposition as the starting state,
\begin{equation}
\langle i|s \rangle = 1/\sqrt{N} ~,~~ 
U_s = H^{\otimes n}(1 - 2|0\rangle\langle 0|)H^{\otimes n} ~~,
\end{equation}
($H$ is the Hadamard operator), the implementation requires
only $O(\log_2 N)$ spatial resources \cite{grover}.
It has been proved that this is the optimal algorithm for
unsorted database search \cite{zalka}.

Several variations of this algorithm have been studied, using different
physical realizations of the database items and the target query oracle.
In the original algorithm, the states are encoded in an $n$-qubit register,
and the oracle is a discrete binary operation (denoted by $U_t$ above).
In the analogue version of the algorithm, the discrete unitary oracle is
traded for a continuous time interaction Hamiltonian (it evolves the target
state somewhat differently than the rest) that acts for the entire duration
of the algorithm, and the number of queries is replaced by the time one has
to wait for before finding the target state \cite{farhi}.
The wave version of the algorithm requires $N$ distinct wave modes,
instead of $n$ qubits, but does not involve quantum entanglement at any
stage \cite{lloyd}.
Such a wave search has also been experimentally implemented using classical
Fourier optics, with a phase-shift plate providing the oracle \cite{spreeuw}.
A classical coupled pendulum model of the analogue version of the
algorithm has been described, where one of the pendulums is slightly
different from the rest and the uniform superposition state $|s\rangle$
is identified with the center-of-mass mode \cite{anirvan}.
In what follows, I describe a binary oracle version of the wave search
algorithm using identical coupled harmonic oscillators.

\section{Harmonic Oscillator Implementation}

A system of coupled harmonic oscillators is frequently studied in physics.
It involves only quadratic forms, and hence can be solved exactly in both
classical and quantum domains. This property makes it extremely  useful in
situations where cross-over between classical and quantum behaviour is to
be analyzed. We shall first look at the classical system, and then observe
that the quantum system essentially follows the same pattern.

\subsection{Classical oscillators}

Let the items in the database be represented by $N$ identical harmonic
oscillators. While they are oscillating in a specific manner, someone taps
one of the oscillators (i.e. elastically reflects it by touching it).
The task is to identify which of the oscillators has been tapped, without
looking at the tapping. The optimization criterion is to design the system
of oscillators, and their initial state, so as to make the identification
as quickly as possible.

\begin{figure}[b]
\epsfxsize=9truecm
\centerline{\epsfbox{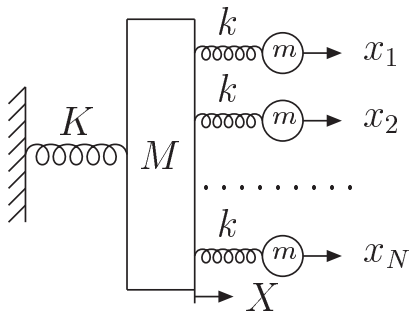}}
\caption{A system of $N$ identical harmonic oscillators,
coupled to a big oscillator via the center-of-mass mode.}
\end{figure}

Grover's algorithm requires identical coupling between any pair of
oscillators. This can be accomplished by coupling all the oscillators
to a big oscillator, as shown in Fig.1. The big oscillator then becomes
an intermediary between any pair of oscillators, with the same strength,
since it is coupled to the center-of-mass mode. The Lagrangian for the
whole system is
\begin{equation}
{\cal L} = \frac{1}{2}M\dot{X}^2 - \frac{1}{2}KX^2
         + \sum_{i=1}^N [ \frac{1}{2}m\dot{x}_i^2 - \frac{1}{2}k(x_i-X)^2 ] ~.
\end{equation}
With the center-of-mass displacement, ${\overline x}\equiv\sum_{i=1}^N x_i/N$,
the Lagrangian can be rewritten as
\begin{eqnarray}
{\cal L} &=& \frac{1}{2}M\dot{X}^2 - \frac{1}{2}KX^2
          +  \frac{1}{2}Nm\dot{\overline x}^2 - \frac{1}{2}Nk({\overline x}-X)^2
\nonumber \\
         &+& \sum_{i=1}^N [ \frac{1}{2}m(\dot{x}_i-\dot{\overline x})^2
          -  \frac{1}{2}k(x_i-{\overline x})^2 ] ~.
\end{eqnarray}

\begin{figure}[b]
\epsfxsize=9truecm
\centerline{\epsfbox{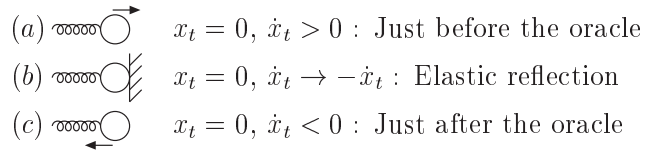}}
\caption{The binary tapping oracle flips the sign of the target oscillator
velocity, when its displacement is zero.}
\end{figure}

Now we can fix the oscillator parameters to implement Grover's algorithm.
In the algorithm, we are interested in the dynamics of the tapped oscillator.
All the other oscillators (i.e. $i \ne t$) influence the dynamics of $x_t$
only through the combination $\overline x$. The dynamics of $(N-2)$ linearly
independent modes orthogonal to $x_t$ and $\overline x$ (they all have the
form $(x_{j \ne t}-x_{k \ne t})$) decouples from the modes of interest;
we can drop them and effectively work in the 3-dimensional space of the
modes $\{X,{\overline x},x_t\}$. (In what follows, we shall first specify
initial conditions such that all $x_{i \ne t}$ are identical and all the
decoupled modes vanish. Subsequently, we will look at the general situation
by adding back all the decoupled modes.)

Choosing units of mass and time such that $m=1,k=1$, and in terms of the
variables
\begin{equation}
Y = \sqrt{M}X ~,~~
{\overline y} = \sqrt{N}{\overline x} ~,~~
y_t = x_t - {\overline x} ~,
\end{equation}
the effective Lagrangian becomes
\begin{eqnarray}
{\cal L}_{\rm eff} &=& \frac{1}{2}\dot{Y}^2 - \frac{1}{2}\frac{K}{M}Y^2
                +  \frac{1}{2}\dot{\overline y}^2
                - \frac{1}{2}\left({\overline y}-\sqrt{\frac{N}{M}}Y\right)^2
\nonumber \\
               &+& \frac{N}{2(N-1)}\dot{y}_t^2 - \frac{N}{2(N-1)}y_t^2 ~.
\end{eqnarray}
The potential energy terms in ${\cal L}_{\rm eff}$ are easily diagonalized,
and yield the eigenvalues
\begin{eqnarray}
&&\omega_\pm^2 = \frac{1}{2}\left(1+\frac{K+N}{M}\right)
  \pm \sqrt{\frac{1}{4}\left(1+\frac{K+N}{M}\right)^2 - \frac{K}{M}} ~,~~
\nonumber \\
&&\omega_+^2 + \omega_-^2 = 1 + \frac{K+N}{M} ~,~~
  \omega_+^2 \omega_-^2 = \frac{K}{M} ~,~~
  \omega_t = 1 ~~.
\end{eqnarray}
The corresponding eigenmodes are
\begin{eqnarray}
e_\pm &=& (1-\omega_\pm^2)Y + \sqrt{\frac{N}{M}}{\overline y}
       =  (1-\omega_\pm^2)\sqrt{M}X + \frac{N}{\sqrt{M}}{\overline x} ~,
\nonumber \\
e_t   &=& y_t = x_t - {\overline x} ~~.
\end{eqnarray}

The initial uniform superposition state can be realized as all the
oscillators moving together, while the big oscillator is at rest.
\begin{equation}
t=0:~~ X=0 ~,~~ \dot{X}=0 ~,~~ x_i = 0 ~,~~ \dot{x}_i = A ~~.
\end{equation}
(We will consider situations with general initial conditions later.)
The reflection operators correspond to shifting the appropriate oscillator
phases by half a period. The binary tapping oracle can be realized as the
elastic reflection illustrated in Fig.2. That implements $U_t$ in the
velocity space, by reversing the target oscillator velocity at the instance
when all the displacements vanish. Time evolution of the coupled oscillators
redistributes the total kinetic energy, and that can implement the operator
$U_s$ with a suitable choice of time interval and frequencies.

With the natural frequency of individual oscillators $\omega=\sqrt{k/m}=1$,
the reflection-in-the-mean operation requires $\omega_\pm$ to be rational
numbers. Optimization means that they should be selected to make the dynamics
of the whole system of oscillators have as small a period as possible.
The solution is not unique. One set of solutions is ($p$ is a positive integer)
\begin{eqnarray}
&& \omega_+ = \frac{2p+1}{2} ~,~~ \omega_- = \frac{1}{2} \nonumber \\
&\Longrightarrow& M = \frac{16Nm}{3(2p+3)(2p-1)} ~,~~
                  K = \frac{(2p+1)^2 Nk}{3(2p+3)(2p-1)} ~~.
\end{eqnarray}
In these cases, in the absence of oracles, the dynamics of the whole system
of oscillators has the period, $T=4\pi$. The big oscillator returns to
its initial rest state ($X=0, \dot{X}=0$), whenever $t$ is an integral
multiple of $2\pi$, i.e. after every half a period. Time evolution for
the same interval of half a period reverses $\dot{\overline x}$, while
leaving $\dot{x}_t - \dot{\overline x}$ unchanged, i.e. it implements
the operator $U_s$ in the velocity space. Thus Grover's algorithm, Eq.(3),
can be realized by applying the tapping oracle at every time interval
$\Delta t=2\pi$.

A more interesting set of solutions is
\begin{equation}
\omega_+ = 2p ~,~ \omega_- = 0
~\Longrightarrow~ M = \frac{Nm}{(2p+1)(2p-1)} ~,~ K = 0 ~.
\end{equation}
In these cases, the big oscillator is not coupled to any support, and
$e_-$ becomes a translation mode for the whole system of oscillators.
The translation mode can be eliminated from the dynamics with the initial
conditions
\begin{equation}
t=0:~~ X=0 ~,~ \dot{X}= -\frac{N}{M}A ~,~ x_i = 0 ~,~ \dot{x}_i = A ~.
\end{equation}
Then, in the absence of oracles, the dynamics of the whole system of
oscillators has the smallest possible period, $T=2\pi$. After half a period,
the big oscillator is back to its initial state, $\dot{\overline x}$ also
returns to its initial value, while $\dot{x}_t - \dot{\overline x}$ changes
its sign. This is equivalent to applying $-U_s$ in the velocity space, and
Grover's algorithm can be implemented by tapping the target oscillator
at every time interval $\Delta t=\pi$.

There is an important physical distinction between the quantum and the
wave interpretations of the amplitude amplification process in Grover's
algorithm---quantum probability is mapped to wave energy. The enhancement
of the quantum amplitude increases the probability of finding the target
state $N$-fold, while the enhancement of the wave amplitude increases the
energy of the target oscillator $N$-fold. The well-known phenomenon of
``beats'' is responsible for energy transfer amongst coupled oscillators.
The elastic reflection oracle does not change energy, and it is interesting
to observe that the oscillator which is obstructed by tapping picks up energy. 

\subsection{Stability considerations}

Now we can look at the behaviour of the wave implementation under general
circumstances. First consider the initial conditions. Despite appearances,
precise synchronization of oscillators is not an issue in the algorithm,
because of the explicit coupling to the center-of-mass mode. For instance,
the algorithm can be started off with an initial push to the big oscillator,
$\dot{X}=B, \dot{x}_i=0$, and the system of oscillators would evolve to the
stage $\dot{X}=0, \dot{x}_i=A$. Furthermore, any arbitrary distribution of
initial velocities of oscillators can be accommodated in the analysis by
bringing back the $(N-2)$ decoupled modes. The decoupled modes have no
effect whatsoever on the dynamics of the $\{X,{\overline x},x_t\}$ modes.
Consequently, the algorithm is only modified to the extent that the energy
amplification of the target oscillator is limited to the initial energy
present in the $\{X,{\overline x},x_t\}$ modes, instead of being $N$-fold.
Explicitly, the maximum gain is
\begin{equation}
\left[ \left( N\dot{\overline x}^2 +
              \frac{N}{N-1}(\dot{x}_t - \dot{\overline x})^2 \right)
       \bigg/ \dot{x}_t^2 \right]_{t=0} ~,
\label{maxgain}
\end{equation}
which can be substantial for the generic situation where the initial
$\dot{x}_t$ and $\dot{\overline x}$ are of the same order of magnitude.

To extract the maximum gain, the algorithm must be stopped at a precise
instant (i.e at a precise value of $Q$); otherwise the evolution continues
in repetitive cycles. The state evolution in Grover's algorithm is a
uniform rotation in the two dimensional $|s\rangle$-$|t\rangle$ subspace.
The average overlap of the target state, with the state $|q(Q)\rangle$
after $Q$ queries, is therefore
\begin{equation}
|\langle q(Q)|t \rangle|^2_{\rm av}
= \langle \sin^2 \theta \rangle_{\rm av} = 1/2 ~.
\end{equation}
Thus if the algorithm is stopped at a random instant, the energy gain
on the average is half of its maximum value in Eq.(\ref{maxgain})---which
can still be substantially larger than $1$.

Next consider the effect of damping \cite{dampingcaveat}.
The crucial ingredient in the algorithm is the coherence amongst the phases
of the oscillators. That is governed by the frequencies of the oscillators,
and is independent of the amplitudes. For a weakly damped oscillator, its
amplitude changes linearly with the damping coefficient, while its frequency
changes quadratically. The time evolution of the above implementation,
therefore, remains essentially unaffected if the oscillators experience a
small damping. The leading effect is a decrease in the energy amplification
due to decaying amplitudes.

Among other variations, simultaneous scaling of masses and spring constants
of the oscillators (i.e. $m_i = \alpha_i m$ and $k_i = \alpha_i k$) does
not alter the algorithm at all, since the scale factors can be absorbed by
redefining $x_i$. One can also contemplate some changes of global conditions
that let the algorithm go through but change its physical interpretation.
For example:
(i) Application of simultaneous tapping oracle to more than one oscillator
(with suitable changes in $M$ and $K$) can focus the energy into the tapped
oscillators.
(ii) Interchange of initial and final states can run the algorithm backwards,
whence the large initial energy of the target oscillator gets uniformly
distributed among all oscillators.
(iii) Elastic reflection of all but the target oscillator implements the
oracle $-U_t$ in the velocity space, in which case it is the unobstructed
oscillator that picks up energy.

\subsection{Quantum domain}

The dynamics of harmonic oscillators is simple enough to permit exact quantum
analysis as well. It is convenient to interpolate between classical and
quantum domains using the coherent state formulation \cite{cohentannoudji}.
Coherent states are superpositions of the eigenstates, parametrized by a
single complex variable $\alpha$,
\begin{equation}
|\alpha\rangle = e^{-|\alpha|^2/2} \sum_n {\alpha^n \over \sqrt{n!}}
               ~ |n\rangle ~.
\end{equation}
They describe Gaussian wavepackets with minimal spread (i.e. displaced
versions of the ground state eigenfunction),
\begin{equation}
\Delta x = \sqrt{\hbar \over 2m\omega} ~,~~
\Delta p = \sqrt{m\hbar\omega \over 2} ~.
\end{equation}
A coherent state with the initial condition $\alpha(t=0)=\alpha_0$ has
energy $\hbar\omega(|\alpha_0|^2+{1\over2})$, and the centre of its
wavepacket performs the same simple harmonic motion as a classical
particle would:
\begin{eqnarray}
\alpha_0 e^{-i\omega t} = {\langle x\rangle(t) \over 2\Delta x}
                        +i{\langle p\rangle(t) \over 2\Delta p} ~.
\end{eqnarray}
The wavefunction of the state evolves according to
\begin{equation}
|\psi(0)\rangle = |\alpha_0\rangle ~~\Longrightarrow~~
|\psi(t)\rangle = e^{-i\omega t/2} |\alpha_0 e^{-i\omega t}\rangle ~,
\end{equation}
while the explicit structure of the wavepacket is given by
\begin{equation}
\psi(x) = \left( {m\omega \over \pi\hbar} \right)^{1/4}
          \exp\left[- \left( x-\langle x\rangle \over2 \Delta x \right)^2
                    + i{x\langle p\rangle \over \hbar} \right] ~.
\end{equation}

\begin{figure}[b]
\epsfxsize=8.5truecm
\centerline{\epsfbox{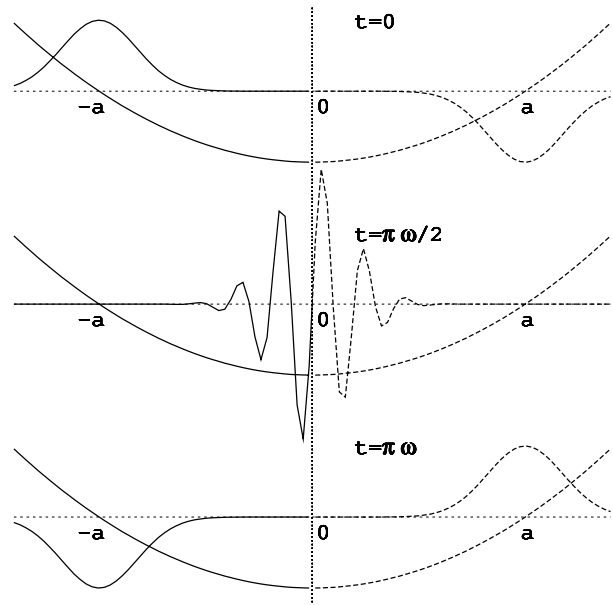}}
\caption{Evolution of the coherent state wavefunction of the tapped
oscillator, with the initial condition $\alpha_0=-a$, The left half of the
figure shows the actual wavepacket in the harmonic oscillator potential,
with the impenetrable wall at $x=0$. The right half of the figure shows
the image wavepacket that ensures the node of the wavefunction at $x=0$.
For $t=\pi\omega/2$, the wavefunction is purely imaginary, but the factor
of $i$ is omitted for convenience in drawing the figure. The wavepacket at
$t=\pi\omega$ includes the geometric phase of $-1$ arising from reflection.}
\end{figure}

Thus the analysis of Section II.A can be carried over unchanged to the
quantum domain, provided we can figure out how the tapping oracle works
for the coherent states. The tapped oscillator corresponds to a particle
moving in the half-oscillator potential
\begin{equation}
V(x) = \textstyle{1\over2} kx^2 ~~{\rm for}~ x\le0 ~,~~
V(x) = \infty ~~{\rm for}~ x>0 ~.
\end{equation}
The impenetrable wall at $x=0$ is equivalent to enforcing the boundary
condition $\psi(x=0)=0$. So the eigenstates of the half-oscillator are
the same as those for the harmonic oscillator, with odd $n$. It is
straightforward to ensure the node at $x=0$ using the method of images,
and the tapped oscillator coherent states become
\begin{equation}
|\alpha_t\rangle = C \big(|\alpha\rangle - |-\alpha\rangle\big) ~,
\end{equation}
with the normalization constant $C=(1-e^{-2|\alpha|^2})^{-1/2}$.
Tapping amounts to change-over between $|\alpha\rangle$ and $|-\alpha\rangle$,
which reverses $\langle x\rangle$ and $\langle p\rangle$ compared to the
untapped motion. In addition, the wavefunction changes sign, which is the
geometric phase corresponding to wave reflection. The evolution of a
coherent state wavepacket undergoing reflection from the wall is depicted
in Fig.3.

\section{Possible Applications}

The oscillator based search process discussed above has the same algorithmic
efficiency as the proposals of Refs.\cite{anirvan,spreeuw,lloyd}---where it
differs from them is in the actual physical implementation. Wave algorithms
are classical, but they have not been explored much historically. The main
reason is that they require exponentially more spatial resources compared to
their digital counterparts, $O(N)$ vs. $O(\log_2 N)$. On the other hand, they
can reduce the number of oracle calls by exploiting superposition of states.
Note that no algorithm based on Boolean logic, either with serial or with
parallel implementation, can reduce the number of oracle calls for an
unsorted database search to less than $O(N)$.

Quantum algorithms are superior to wave algorithms, because they can use
superposition as well as reduce spatial resources. The reduction of spatial
resources, however, comes with the cost that quantum algorithms have to
work with entangled states. Quantum entanglement is far more sensitive to
decoherence caused by environmental disturbances than mere superposition,
and that has made physical implementations of quantum algorithms very
difficult. On the other hand, superposition of classical waves can be
fairly stable, even when a small amount of damping is present, and that
can make wave implementations advantageous in specific physical contexts.

These comparisons suggest that wave algorithms fall in a regime inbetween
classical and quantum algorithms---more efficient than the former and more
robust than the latter. They are likely to be useful in practical situations,
where $N$ is not very large and environmental disturbances are not negligible.
Indeed it is worthwhile to systematically explore them, just like randomized
algorithms have been \cite{motwani}.

In the specific case of the unsorted database search problem, the remarkable
simplicity of the oscillator implementation makes one wonder about possible
applications, besides constructing convenient demonstration models.
The implementation of Ref.\cite{anirvan} has one oscillator with a frequency
different from the rest, which gets singled out by dynamical evolution of
the coupled system. In the present scenario, all oscillators are identical,
but one of them is discretely tapped by an external agency and gets selected.
I point out a physical situation below, that can fit such a scenario,
and where involvement of new mechanisms can enhance our understanding of
the observed phenomena.

\subsection{Catalysis}

The practically useful property of the wave search algorithm is that it
focuses energy into one of the oscillator modes. There exist a large
number of chemical reactions which, although not forbidden by energy
conservation, are extremely slow because they have to pass through an
intermediate state of high energy. In these reactions, the dominant term
governing the reaction rate is the Boltzmann factor, $\exp(-E_b/kT)$, with
the barrier energy in the exponent. Only a tiny fraction of the molecules
in the tail of the thermal distribution are energetic enough to go over
the barrier and complete the reaction. It is known that the rates of many
such reactions can be enhanced by orders of magnitude by adding suitable
catalysts (enzymes in case of biochemical reactions) to the reactants.
The conventional explanation for the reaction rate enhancement, called
transition state theory, is that the catalysts lower the energy of the
intermediate state by modifying the chemical environment of the reactants.

The preceding analysis of the wave search algorithm suggests another
mechanism for catalysis. Vibrations and rotations of molecules are
ubiquitous harmonic oscillator modes. The catalyst can act as the big
oscillator and focus energy of many modes into the reactant which faces the
energy barrier. For example, the catalytic substrate can have many identical
molecules of one reactant stuck to it and vibrating, the second reactant
then comes drifting along and interacts with one of the stuck molecules,
that molecule picks up energy from its neighbours and the reaction gets
completed. In such a scenario, for maximum efficiency, the physical
parameters (masses and spring constants) need to have specific values.
But even without perfectly tuned parameters, there can be partial
energy focusing that provides useful increase in the reaction rate.
Whether this mechanism exists among the known catalysts, or whether we can
design new type of catalysts that use this mechanism, is an open question.

The catalytic role of chemical environment vs. physical waves can be tested
by isotopic substitution in the reactants, since isotopic substitution
changes physical parameters without altering chemical properties.
The conventional transition state theory has no mass dependence,
so any isotope dependence of reaction rates is a signal of involvement
of physical (in contrast to chemical) features in the process.
Many examples of isotopic dependence of catalytic reaction rates have
been discovered, and the effect is referred to as the ``Westheimer effect''
or the ``kinetic isotope effect'' \cite{westheimer}.
The effect is the largest for substitution of hydrogen by deuterium,
and has been extensively studied for the rupture of C-H/C-D bonds.
The transition state theory has been found inadequate for theoretical
understanding of the observations, and vibrationally enhanced quantum
tunneling has been invoked as an alternative \cite{VEGST}.

In this context, the oscillator based search process described above has
two novel features to suggest yet another alternative:\\
(i) The energy focusing mechanism enhances reaction rates beyond their
naive classical values. This enhancement is both mass and temperature
dependent, because the energy enhancement depends on the masses involved
while the Boltzmann factor contains the temperature.\\
(ii) The wavefunction sign-flip caused by reflection can switch between
bonding and anti-bonding molecular orbitals, and thus help in the transfer
of atoms. This feature is not related to either mass or temperature.\\
It would be therefore worthwhile to explore the oscillator inspired
catalytic mechanism with careful modeling of specific reactions.

In closing, I mention an intriguing possibility. I had constructed a quantum
database search scenario for the DNA replication process, whose stumbling
block was the maintenance of quantum coherence in presence of continuous
jostling of molecules \cite{patel1}.
It should be feasible to construct a wave database search scenario involving
vibrational and rotational modes of the molecules. That would be far less
troubled by decoherence effects, and make interpretation of the genetic
language as optimal solution to the database search process likely.


\begin{thebibliography}{99}
\bibitem{dampingcaveat}{Damping for a harmonic oscillator has to be
         analyzed using its equation of motion; it cannot be described
         using a time-independent Lagrangian formulation.}
\end{thebibliography}
\end{document}